\documentclass[epsf,11pt]{article}
\usepackage{epsf}
\usepackage{amsmath}
\usepackage{amssymb}
\usepackage{graphicx}

\title{ Reduction of su(N) loop tensors to trees}
\author{Maciej Trzetrzelewski \footnote{trzetrzelewski@th.if.uj.edu.pl} \\
M. Smoluchowski Institute of Physics, Jagiellonian University \\
Reymonta 4, 30-059 Cracow, Poland \\
  }
\begin{document}
\maketitle \abstract{ We present a systematic method to express all $su(N)$ invariant tensors in terms of forests i.e. products of tree tensors. }

\section{Introduction}

It is a well known fact that in a simple  Lie algebra of rank $r$
there are exactly r independent Casimir invariants. This causes a
direct restriction on algebra of su(N) tensors since naively one
could produce an infinite amount of invariants by contracting d,f
tensors. The Cayley-Hamilton theorem which is the reason of why higher Casimir invariants are dependant on the first r ones clearly gives
additional $d_{ijk}$ tensor identities. The systematic,
computer friendly, approach to obtain these formulas  was presented by
Sudbery [1].  One may also define
matrices  $[F_i]_{jk}=f_{ijk}$, $F=a_iF_i$, $a_i \in \mathbf{C}$
and use Cayley-Hamiton equation to obtain analogous identities for $f_{ijk}$ which
was elaborated in details in [2]. In this paper we will use a geometrical approach to find formulas  on $su(N)$ loop tensors in term of $su(N)$ tree tensors. In this way we give
a recursive method which allows to express any su(N) invariant
tensor in terms of basic ones i.e. forests ( products of trees ).
 In section 3 we prove several lemmas and eventually the main result. 
 In section 4 we present a few examples to give the insight into the method.

We will use the following conventions

\begin{equation}
\lambda_i\lambda_j=\frac{2}{N}\delta_{ij}\mathbf{1}+d_{ijk}\lambda_k+if_{ijk}\lambda_k,
\end{equation}

\noindent where $\lambda_i$'s are $su(N)$   generators in
fundamental representation  and $d_{ijk}$,
$f_{ijk}$  are complectly symmetric/antisymmetric structure
tensors. Multiplication law (1) together with Jacobi identities
for $\lambda_i$'s give identities known long time ago [3]. We will
make a special use of

\begin{equation}
f_{i_1 i_2 k}d_{k i_3 i_4 }+f_{i_1 i_3 k}d_{k i_2 i_4 }+f_{i_1 i_4
k}d_{k i_2 i_3 }=0,
\end{equation}

\noindent and

\begin{equation}
f_{i_1 i_2 k}f_{k i_3 i_4 }=\frac{2}{N}\left(\delta_{i_1 i_3
}\delta_{i_2 i_4}-\delta_{i_1 i_4 }\delta_{i_2 i_3} \right)+d_{i_1
i_3 k}d_{k i_2 i_4 }-d_{i_1 i_4 k}d_{k i_2 i_3 }.
\end{equation}

\section{Bird tracks }

In order to grasp the variety of all  possible invariant tensors
it is helpful to introduce the diagrammatic notation for d and f
tensors (figure 1). Each leg corresponds to one index and summing
over any two indices is simply gluing appropriate legs.  This
notation is very convenient because now any tensor may be
represented by a graph.

\begin{figure}[h]
\centering \leavevmode
\includegraphics[width=0.6\textwidth]{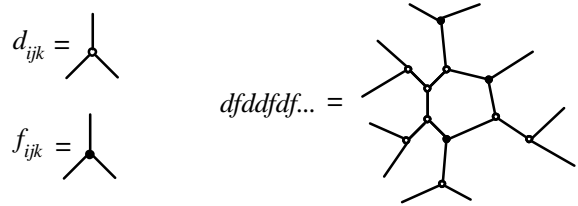}
\caption{$d_{ijk}$, $f_{ijk}$ diagrams and a typical tensor
diagram.}
\end{figure}

Such diagrammatic approach has already been introduced long time
ago by Cvitanovi\v{c}. The d, f tensors are called bird tracks since they
look like tracks of a bird. The reader is referred to [4] where a
vast amount of group properties is rediscovered in such
diagrammatic language. Since $d_{ijk}$ is totaly symmetric the
order of corresponding legs is irrelevant. For $f_{ijk}$ we have
to set e.g. anticlockwise convention. A special group of diagrams
are loop and tree diagrams (figure 2)

\begin{figure}[h]
\centering \leavevmode
\includegraphics[width=0.4\textwidth]{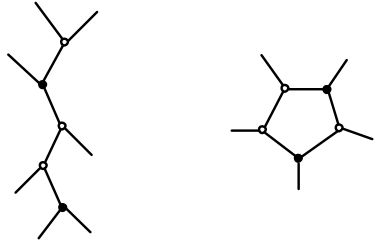}
\caption{A tree and a loop diagram.}
\end{figure}

\noindent One may rotate any diagram on the  plane without
changing the value of the corresponding tensor. Reflections ( or
rotations in three dimensions ) are allowed as well, however in
this case one has to take care of the sign since f tensor is
antisymmetric. If a diagram consists only of d tensors then
reflections will not affect its value. Several definitions are now
in order.

The index that corresponds to d/f tensor is  called d/f index. A
loop $L_1$ is smaller then loop $L_2$ if the number of f,d tensors
in $L_2$ is smaller then the number of f,d tensors on $L_1$. Note
that, in general, trees can be attached to loops. In that case we
will call it a tree loop diagram. Similarly a tree loop $L_1$ is
smaller then tree loop $L_2$ if the number of f,d tensors within
loop in $L_2$ is smaller then the number of f,d tensors within
loop in $L_1$.

A loop diagram is called $n$ loop if it  consists of n tensors. A
loop diagram is called $d$ loop if it consists of d tensors only.
A loop diagram is called $1f$/$2f$ loop if it consists of one/two
f tensor and d tensors.

\section{Loop reduction}

This section consists of several lemmas  and eventually a theorem
which gives a computational method for expressing loops by trees.

\vspace{0.5cm}

 \textbf{Lemma 1}. Any loop is a linear combination of d loops and 1f loops.

\vspace{0.5cm}

 \textit{Proof.} Let us rewrite (3) in diagrammatic notation \footnote{Instead of writing $i_1,i_2,i_3,i_4$ ect. we prefer $1,2,3,4$ since it causes no misunderstanding and gives a better idea of the structure of indices. Take attention of the order of indices 3 and 4. There is no mistake. The whole diagram is supposed to be read with the anticlockwise convention. }

\begin{figure}[h]
\centering \leavevmode
\includegraphics[width=0.5\textwidth]{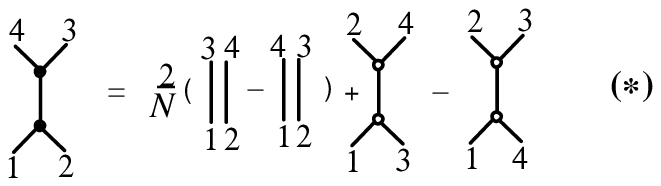}
\end{figure}

\noindent therefore it is sufficient to consider loops where f tensor is between d tensors

\begin{figure}[h]
\centering \leavevmode
\includegraphics[width=0.15\textwidth]{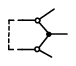}
\end{figure}

\noindent However in such case one can use Jacobi identities (2)

\begin{figure}[h]
\centering \leavevmode
\includegraphics[width=0.7\textwidth]{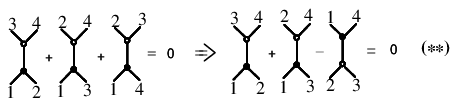}
\end{figure}

\pagebreak

\noindent therefore attaching \footnote{The question mark means that there may be f tensor or d tensor.}

\begin{figure}[h]
\centering \leavevmode
\includegraphics[width=0.15\textwidth]{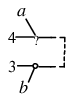}
\end{figure}

\noindent we get

\begin{figure}[h]
\centering \leavevmode
\includegraphics[width=0.7\textwidth]{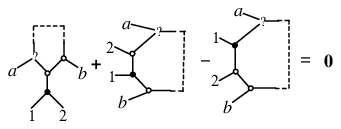}
\end{figure}

\noindent The last identity means that we can  "move" f tensor
along the loop producing a smaller tree loop. Eventually such f
tensor will "meet" another f tensor ( if there is another one in
the loop ) and one can use $(\ast)$ again to get rid of f tensors.
This procedure stops on d loops or 1f loops.$\square$

\vspace{0.5cm}

\pagebreak

 \textbf{Lemma 2}. Let A be a d loop or 1f loop. Then any permutation of d indices of A does not change the value of A up to trees and smaller tree loops.

\vspace{0.5cm}

 \textit{Proof.}  Consider identity $(\ast)$ and attach the tensor

\begin{figure}[h]
\centering \leavevmode
\includegraphics[width=0.2\textwidth]{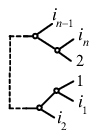}
\end{figure}

\noindent where the dashed lines correspond to d tensors only.

 \noindent The result is

\begin{figure}[h]
 \centering \leavevmode
\includegraphics[width=1.2\textwidth]{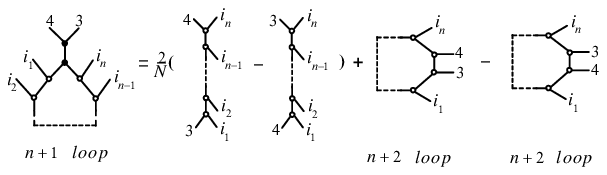}
\end{figure}

\noindent Therefore the permutation of  indices $3,4$ does not
change the d loop up to trees and smaller tree loops. Since
indices $3,4$ are not distinguished we can do any permutation of
any two indices and the d loop will not change the value up to
trees and smaller tree loops. Since any permutation is a proper
composition of transpositions the Lemma 2 follows for d loops. The
proof for 1f loops is analogous.$\square$

\vspace{0.5cm}

\pagebreak

 \textbf{Lemma 3}. Any 1f loop is a linear combination of trees and smaller tree loops.

 \vspace{0.5cm}

 \textit{Proof.} Consider the Jacobi identities $(\ast\ast)$ and attach the tree diagram ( consisting of d tensors only )

\begin{figure}[h]
\centering \leavevmode
\includegraphics[width=0.3\textwidth]{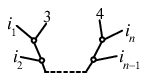}
\end{figure}

\noindent The result is

\begin{figure}[h]
\centering \leavevmode
\includegraphics[width=0.8\textwidth]{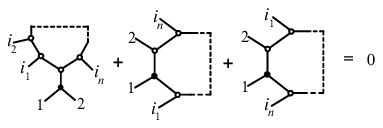}
\end{figure}

\noindent From Lemma 2 it follows that

\begin{figure}[h]
\centering \leavevmode
\includegraphics[width=1\textwidth]{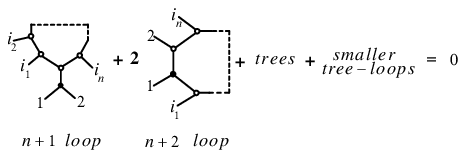}
\end{figure}

\vspace{0.5cm}

\pagebreak

 \textbf{Lemma 4}. Any d loop is a linear combination of trees and smaller tree loops.

 \vspace{0.5cm}

 \textit{Proof.} Consider Jacobi identity $(\ast\ast)$ and attach the following tree ( consisting of one f tensor and
$n-1$ d tensors )

\begin{figure}[h]
\centering \leavevmode
\includegraphics[width=0.2\textwidth]{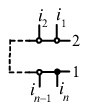}
\end{figure}

\noindent The result is

\begin{figure}[h]
\centering \leavevmode
\includegraphics[width=0.9\textwidth]{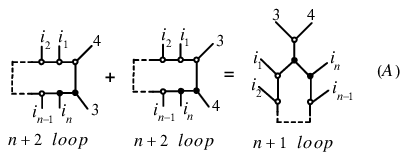}
\end{figure}

\noindent Therefore the symmetrization of indices 3,4 in such 2f loop is equal to smaller tree loop. Now for the proof of Lemma 4 consider  identity $(\ast)$ and
attach the tree tensor ( consisting of d tensors only )

\begin{figure}[h]
\centering \leavevmode
\includegraphics[width=0.2\textwidth]{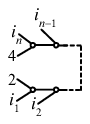}
\end{figure}

\pagebreak

\noindent The result is

\begin{figure}[h]
\centering \leavevmode
\includegraphics[width=1.2\textwidth]{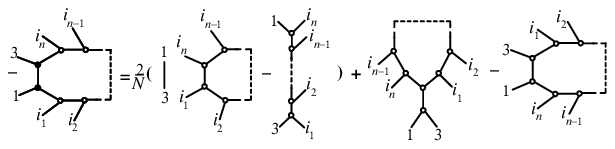}
\end{figure}

\noindent Or simply

\begin{figure}[h]
\centering \leavevmode
\includegraphics[width=1\textwidth]{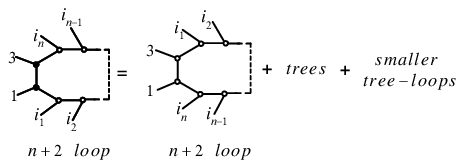}
\end{figure}

\noindent According to $(A)$ the symmetrization over indices 1 and $i_1$ gives

\begin{figure}[h]
\centering \leavevmode
\includegraphics[width=1\textwidth]{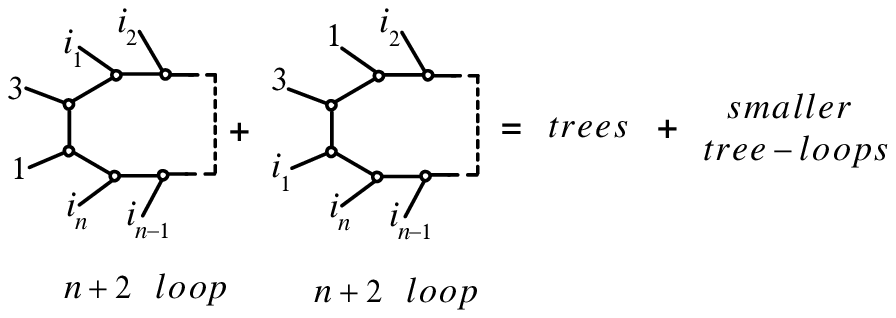}
\end{figure}

\noindent Due to Lemma 2 the Lemma 4 follows.$\square$

\vspace{0.5cm}

\pagebreak

 \textbf{Theorem}. Any loop diagram is a linear combination of forests.

\vspace{0.5cm}

 \textit{Proof.} From Lemma 1 it is sufficient to consider d loops and 1f loops. From Lemma 3 and Lemma 4 we may recursively reduce 1f loop and d loop to arbitrary small loops and ultimately to trees. $\square$

\vspace{0.5cm}

 \textbf{Corollary 1}. Any diagram is a linear combination of forests.

\vspace{0.5cm}

\textit{Proof.} Any loop in the diagram  may be replaced by a
linear combination of trees. This will in general produce more
loops however the number of d,f tensors will be smaller after such
replacement. Following the induction with respect to the number of
d,f tensors we finely reduce all loops. $\square$

\vspace{0.5cm}

\textbf{Corollary 2}. Any diagram is a linear  combination of
products of trace tensors  $Tr(\lambda_{i_1} \ldots \lambda_{i_n})$ where
$\lambda_i$'s are $su(N)$ Gell-Mann matrices.

\vspace{0.5cm}

\textit{Proof.}  According to Corollary 1 it is  sufficient to
consider tree diagrams. With help of (1) we have

\[
Tr(\lambda_{i_1}\ldots \lambda_{i_n})=\frac{1}{N}Tr(\lambda_{i_1}\lambda_{i_2})Tr(\lambda_{i_3}\ldots \lambda_{i_n})+(d_{i_1 i_2 k}+if_{i_1 i_2 k})Tr(\lambda_k \lambda_{i_3}\ldots \lambda_{i_n})
\]

\noindent therefore
\[
d_{i_1 i_2 k}Tr(\lambda_k \lambda_{i_3}\ldots \lambda_{i_n})=\frac{1}{2} Tr(\lambda_{(i_1}\lambda_{i_2)}\lambda_{i_3}\ldots \lambda_{i_n})
-\frac{1}{N}Tr(\lambda_{i_1}\lambda_{i_2})Tr(\lambda_{i_3}\ldots \lambda_{i_n})
\]
\noindent and
\[
f_{i_1 i_2 k}Tr(\lambda_k \lambda_{i_3}\ldots \lambda_{i_n})=\frac{1}{2} Tr(\lambda_{[i_1}\lambda_{i_2]}\lambda_{i_3}\ldots \lambda_{i_n})
\]

\noindent  hance Corollary 2 follows by induction . $\square$

\section{Examples}

Below we give $su(N)$ formulae for the lowest d loops i.e.
triangles, squares and pentagons. The identities for triangles and
squares are already in the literature in [3] and [5] respectively.
However to the knowledge of the author these identities are
missing for pentagons and higher loops. The results are

\begin{figure}[h]
\centering \leavevmode
\includegraphics[width=0.4\textwidth]{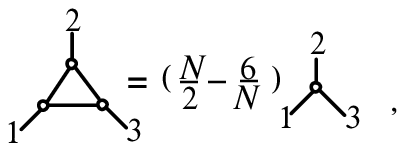}
\end{figure}

\begin{figure}[h]
\centering \leavevmode
\includegraphics[width=0.9\textwidth]{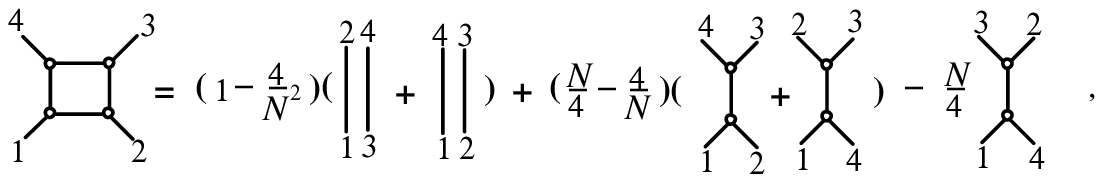}
\end{figure}

\begin{figure}[h]
\centering \leavevmode
\includegraphics[width=1\textwidth]{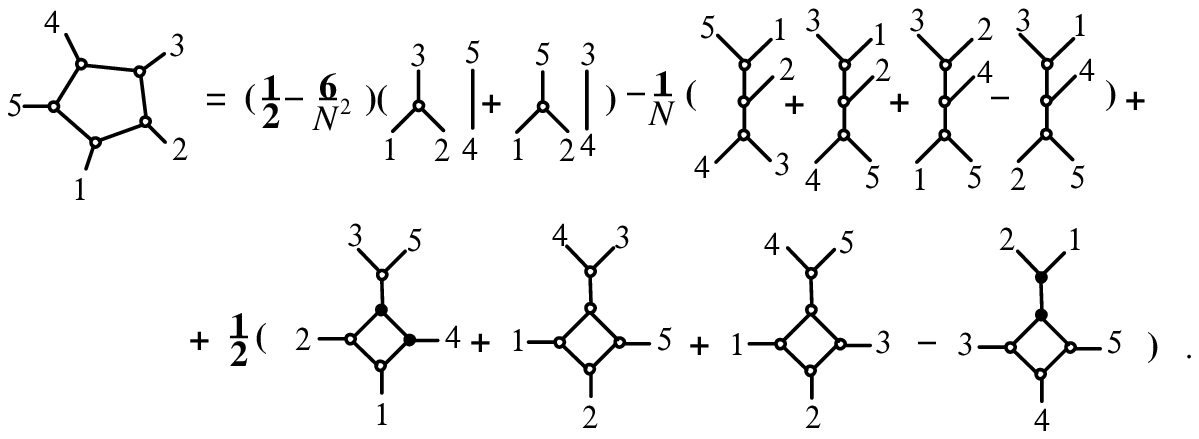}
\end{figure}
\noindent The last identity in standard notation is
\[
Tr(D_{i_1}D_{i_2}D_{i_3}D_{i_4}D_{i_5})=(\frac{1}{2}-\frac{6}{N})(d_{i_1 i_2 i_3} \delta_{i_4 i_5}+d_{i_1 i_2 i_5} \delta_{i_4 i_3}    )+
\]
\[
-\frac{1}{N}(d_{i_3 i_4 k}d_{k i_2 l}d_{l i_1 i_5}+d_{i_4 i_5 k}d_{k i_2 l}d_{l i_1 i_3}+d_{i_1 i_5 k}d_{k i_4 l}d_{l i_2 i_3}-d_{i_2 i_5 k}d_{k i_4 l}d_{l i_1 i_3}    )+
\]
\[
+\frac{1}{2}( Tr(D_{i_2}D_{i_1}F_{i_4}F_{k})d_{k i_5 i_3}+Tr(D_{i_1}D_{i_2}D_{i_5}D_{k})d_{k i_3 i_4} +Tr(D_{i_1}D_{i_2}D_{i_3}D_{k})d_{k i_5 i_4} -Tr(D_{i_3}D_{i_4}D_{i_5}F_{k})f_{k i_1 i_2}     )
\]
 It should be noted that all these identities have been verified in Mathematica with perfect agreement.
\section{Summary}

The aim of this paper was to give a systematic approach to compute
loop tensors. The reason of doing so lies in the analysis of
systems based on $su(N)$ group. In fact the author came across
this problem while studying supersymmetric Yang-Mills quantum
mechanics for arbitrary $N$ and large $N$ limit [7,8]. This issues will
be published elsewhere. The method agrees with recent results [2]
where the problem was solved via characteristic equation for F
matrices. Let us note that it is a laborious task to obtain
this equation for arbitrary $su(N)$ therefore a big loop diagram
for large N is in general difficult to reduce. In diagrammatic
approach this problem does not exist since we make no use of
characteristic equation. Indeed lemmas presented here are so
simple that one  could write a computer program for arbitrary loop
reduction. What is even more remarkable is that the derivation of our result is based only on Jacobi identities and multiplication law (1). We did not use the relations derived by Sudbery [1] although it is evident
that one may contract his formulas with eg. $d_{ijk}$ providing a constraint on a d loop.

The diagrammatic method may be applied to arbitrary Lie algebra.
However since the multiplication rule (1) is different in other
cases than $su(N)$ we expect the conclusions to be deferent.
Indeed in $g_2$ case the situation is so different that the
simplest triangle d loop is not proportional to $d_{ijk}$ tensor
[6].

Finely let us note that the method [2] gives no information about lower
degree traces (eg. $Tr(F^4)$,$Tr(F^6)$,$Tr(F^8)$,$Tr(F^{10})$ in
$su(5)$ cannot be written as polynomials in lower degree traces).
One may however apply different arguments [5] to derive formulae
for four-fold traces. Our results also agree with them. Unfortunately
these arguments get more complicated while analyzing bigger loops.

\section{Acknowledgments}
I thank J. Wosiek and A. Macfarlane for discussions and
encouragement. This work was supported by the Polish Committee for
Scientific Research under grant no. PB 1P03B 02427 (2004-2007).

\end{document}